\vspace{-20pt}
\documentclass[9.96pt,journal]{IEEEtran}
\vspace{-20pt}
\usepackage{balance}
\ifCLASSINFOpdf
\else
\usepackage[dvips]{graphicx}
\fi
\usepackage{setspace}
\usepackage{cite}
\usepackage{bm}
\usepackage{mathrsfs}
\usepackage{algorithmic}
\usepackage{algorithm}
\usepackage{graphicx}
\usepackage{psfrag}
\usepackage{subfigure}
\usepackage{url}
\usepackage{stfloats}
\usepackage{amsmath,amssymb,amsfonts}
\usepackage{empheq}
\usepackage{amssymb}
\usepackage{array}
\usepackage{balance}
\usepackage{lipsum}
\usepackage{multicol}
\usepackage{epstopdf}
\usepackage{caption}
\usepackage{geometry}
\usepackage{graphicx}
\usepackage{makecell,multirow,diagbox}
\usepackage{textcomp,booktabs}
\usepackage[usenames,dvipsnames]{color}
\usepackage{colortbl}
\definecolor{mygray}{gray}{.9}
\definecolor{mypink}{rgb}{.99,.91,.95}
\definecolor{mycyan}{cmyk}{.3,0,0,0}

\ifCLASSOPTIONcompsoc
\usepackage[caption=true, font=footnotesize, labelfont=sf, textfont=sf]{subfig}
\else
\usepackage[caption=true, font=footnotesize]{subfig}
\begin{document}
	
	\title{\vspace{-0.5em}\LARGE Robust Trajectory and Communication Design in IRS-Assisted UAV Communication under Malicious Jamming \\
	\thanks{ 
 (\textit{Corresponding{\text{ }}author:{\text{ }}Wendong{\text{ }}Yang.} email: ywd1110@163.com) 
	}
}	

	\author{\IEEEauthorblockN{Zhi Ji\IEEEauthorrefmark{1},
	    Xinrong Guan\IEEEauthorrefmark{1},
	    Jia Tu\IEEEauthorrefmark{2},
	    Qingqing Wu\IEEEauthorrefmark{3},
		and Wendong Yang\IEEEauthorrefmark{1}}\\
	\IEEEauthorblockA{\textit{\IEEEauthorrefmark{1}College of Communications Engineering, Army Engineering University of PLA, Nanjing, China,\\ \IEEEauthorrefmark{2}College of International Studies, National University of Defense Technology, Nanjing, China,\\
			  \IEEEauthorrefmark{3}State Key Laboratory of Internet of Things for Smart City, University of Macau, Macau, China} \\
		}
	
}	
	
\maketitle
	
\begin{abstract}
In this paper, we study an unmanned aerial vehicle (UAV) communication system, where a ground node (GN) communicate with a UAV assisted by intelligent reflecting surface (IRS) in the presence of a jammer with imperfect location information. We aim to improve the achievable average rate via the joint robust design of UAV trajectory, IRS passive beamforming and GN’s power allocation. However, the formulated optimization problem is challenging to solve due to its non-convexity and coupled variables. To overcome the difficulty, we propose an alternating optimization (AO) based algorithm to solve it sub-optimally by leveraging semidefinite relaxation (SDR), successive convex approximation (SCA), and S-procedure methods. Simulation results show that by deploying the IRS near the GN, our proposed algorithm always improves the uplink achievable average rate significantly compared with the benchmark algorithms, while deploying the IRS nearby the jammer is effective only when the jammer's location is perfectly known. 

\end{abstract}


\section{Introduction}
		\label{Introduction}
Benefiting from the flexible mobility and deployment, unmanned aerial vehicles (UAVs) have been employed in numerous applications in recent years, such as remote surveillance, photography,  traffic control, and cargo transportation, etc \cite{1}. By the reasonable design of its trajectory, UAV tends to obtain more dominant line-of-sight (LoS) channels and thus greatly improves the communication performance \cite{2}. However, these advantages are confronted with many challenges. Due to the broadcast nature of wireless transmission and the strong LoS links,  UAVs are more vulnerable to attacks from jamming and eavesdropping \cite{3}.
	
Due to its capability of smartly configuring the wireless propagation environment, intelligent reflecting surface (IRS) has emerged as a promising technology for future wireless networks \cite{5,huang1}. Specifically, IRS is a uniform planar array consisting a large number of passive reflecting units, which can reflect the wireless signal with adjustable phase shift and/or amplitude. The main advantage of IRS-assisted wireless communications lies in that it can considerably increase the channel capacity with low power consumption and high deployment flexibility. Benefiting from above superiority, IRS has been applied to cognitive radio \cite{cog2}, device-to-device (D2D) networks \cite{8}, secure wireless communication \cite{11} and so on. A comprehensive tutorial on IRS-aided wireless communications can be found in \cite{12}.

Thanks to the appealing advantages, IRS has been exploited as a promising choice for improving the performance of UAV communication for combating security threats. In \cite{17}, the average secrecy rate was maximized by jointly optimizing the UAV trajectory, the transmit beamforming and the phase shift of IRS. In \cite{ICC} and \cite{WCL}, the UAV equipped with one IRS serves as a passive relay is proposed to maximize the secrecy rate of the system.
By exploiting the IRS, higher average rate of the UAV communication has been achieved to mitigate the jamming signal from the malicious jammer \cite{21}. However, all of these works are based on the perfect knowledge of the eavesdropper/jammer's location, which is generally unavailable in practical systems and thus brings new challenges to the IRS passive beamforming, deployment, and UAV trajectory. To the best of our knowledge, this still remains an open problem and needs further study.
\begin{figure}
	\centering
	\includegraphics[width=8.0cm]{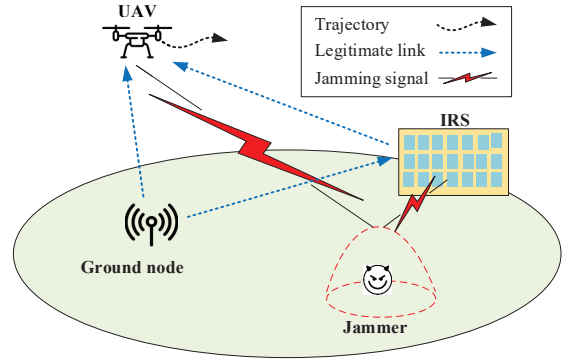}
	\caption{IRS-assisted UAV communication in the presence of a jammer.} \label{Model}
		\vspace{-5mm}
\end{figure}
	
In this paper, we study a UAV communication system aided by IRS, and the communication process incurs malicious jamming from a jammer with imperfect location information, as shown in Fig. \ref{Model}. We consider a joint robust design to maximum the achievable average rate in the uplink. However, the formulated optimization problem is difficult to tackle due to its non-convexity and coupled variables. To deal with the problem, we propose an alternating algorithm by using the successive convex approximation (SCA), semidefinite relaxation (SDR), and S-procedure methods to require the sub-optimal solution. For the case that IRS is deployed near the GN, numerical results show that our proposed robust algorithm can greatly improve the worst system performance compared to the benchmark algorithms.
However, for the case that IRS is deployed near the jammer, numerical results show that the proposed algorithm is effective only with perfect jammer's location informantion.

\textit{Notations}: ${\text{diag}}\left\{  \cdot  \right\}$, ${\text{tr}}\left(  \cdot  \right)$ returns the diagonal matrix whose diagonals are the elements of input vector. $\Re \left(  \cdot  \right)$ and $\angle  \left(  \cdot  \right)$ represent the real part and the phase of the input complex value. By denoting ${\mathbf{\Theta}}\underline  \succ   0$, it is used to represent ${\mathbf{\Theta }}$ as positive semidefinite.
	
\section{System Model and Problem Formulation}
\label{sec:System Model}
	
In this paper, a UAV communication system is considered as shown in Fig.\ref{Model}, where an IRS is deployed to assist in the information transmission from the GN to a UAV in the presence of a jammer with imperfect location information. All communication nodes are placed in the three dimensional (3D) Cartesian coordinates. The location of the jammer, GN and IRS are expressed as ${\bf{q_m}} = [{x_m},{y_m},{z_m}]$, ${\bf{q_g}} = [{x_g},{y_g},0]$ and ${\bf{q_r}} = [{x_r},{y_r},{z_r}]$, respectively. Due to the uncertain location region in 3D space, we regard the jammer’s location as a hemisphere, which roughly simulates the practical situation. In the result, the location of the jammer can be denoted by
\begin{equation}\label{eq0}
	\left\{ {\begin{array}{*{20}{c}}
		{{x_m} = {\bar x_{m}} + \Delta {x_m},} \\ 
		{{y_m} = {\bar y_{m}} + \Delta {y_m},} \\ 
		{{z_m} = {\bar z_{m}} + \Delta {z_m},} 
	\end{array}} \right.
\end{equation}
where ${\bar q_{m}} = [{\bar x_{m}},{\bar y_{m}},{\bar z_{m}}]$ 
denotes the center of hemisphere and $(\Delta {x_m}, \Delta {y_m}, \Delta {z_m}) \in {\varepsilon _m}$ denotes the possible estimated errors, which is confined to the condition 
\begin{equation}\label{eq0}
	{\varepsilon _m} \triangleq \{ \Delta x_m^2 + \Delta y_m^2 + \Delta z_m^2 \leqslant D_m^2,\Delta z \geqslant 0\}, 
\end{equation}
where ${D_m}$ is the distance deviated from the center of the hemisphere.
	
The UAV is assumed to fly at a fixed height $H_u$. The flying time of the UAV is $T$, which is divided into $N$ time slots, i.e., $\Delta t = T/N$, where ${\Delta t}$ is the length of a time slot. The UAV has the fixed starting point and destination, which are denoted by ${{\mathbf{q}}_{0}}$ and ${{\mathbf{q}}_{N}}$. The trajectory of the UAV can be expressed as ${\bf q}[n] = {[x[n],y[n],z[n]]^T},n \in {\cal{N}} = \{ 1,2,...,N\}$, ${\bf{Q}} \buildrel \Delta \over = \{ {\bf q}[n],\forall n\}$, which meets the mobility constraints as 
\begin{equation}\label{eq0}
	{\bf{q}}\left[ 0 \right] = {{\bf{q}}_{0}},{\bf{q}}\left[ N \right] = {{\bf{q}}_{N}},\\
\end{equation}
\begin{equation}\label{eq0}
	\left\| {{\mathbf{q}}[n] - {\mathbf{q}}[n - 1]} \right\| \leqslant {D_{\max }},n = 1,...,N,\\
\end{equation}
where ${{D_{\max }}}$ is the maximum flying length in each time slot. $p[n]$ is the transmit power of the GN in time slot $n$ and ${\mathbf{P}} = \left\{ {p\left[ n \right],\forall n} \right\}$, thus the power constraints are expressed as	
\begin{equation}\label{eq0}
	\frac{1}{N}\sum\limits_{n = 1}^N {p[n]}  \leqslant \bar p,\\~~
	p[n] \leqslant {p_{{\text{max}}}},\forall n,
\end{equation}
where $\bar p$ and ${p_{max}}$ are the average transmit power and the maximum transmit power of the GN, respectively.
We assume that both the GN and the UAV are equipped with single omni-directional antenna, and the IRS composes of a uniform planar array (UPA) with $L=L_x \times L_z$ elements, where $L_x$ and $L_z$ denote the number of elements along the x-axis and z-
axis, respectively. Then we denote the diagonal phase-shift matrix for the IRS as ${\mathbf{\Gamma }} \triangleq \left\{ {{\mathbf{\Gamma }}[n] = {\text{diag}}\left( {{e^{j{\theta _1}[n]}},{e^{j{\theta _2}[n]}},...,{e^{j{\theta _L}[n]}}} \right),\forall n} \right\}$, where ${\theta _i}[n] \in [0,2\pi ),i \in \{ 1,...,L\}$ is the phase shift of the $i$-th reflecting element in time slot $n$.
Due to high flying altitude of UAV and the flexible deployment of IRS, we assume that all channels are LoS channels in the considered system. Specifically, the GN-UAV channel in time slot $n$ is expressed as
\begin{equation}\small
	{{\text{h}}_{gu}}[n] = \sqrt {\rho d_{gu}^{ - 2}\left[ n \right]} {e^{{{ - j2\pi {d_{gu}}\left[ n \right]} \mathord{\left/
	{\vphantom {{ - j2\pi {d_{gu}}\left[ n \right]} \lambda }} \right.
	\kern-\nulldelimiterspace} \lambda }}},
\end{equation}
where ${d_{gu}}\left[ n \right] = \left\| {{\bf{q}}\left[ n \right] - {{\bf{q}}_G}} \right\|$ is the distance between the GN and the UAV. $\lambda$ is the carrier wavelength and $\rho$ is the path loss at the reference distance ${D_0} = 1{\rm{m}}$. The same channel model is adopted for the channel from the jammer to the UAV, i.e., ${h_{mu}[n]}$. 
Further, the reflecting channel such as GN-IRS-UAV channel is composed of two parts, namely, the GN-IRS channel and the IRS-UAV channel. Specifically, the IRS-UAV channel denoted by ${{\bf{h}}_{ru}}\left[ n \right] \in {\mathbb{C}}^{L \times 1}$, can be given by 
\begin{equation}\small
    {{\mathbf{h}}_{ru}}[n] = \sqrt {\rho d_{ru}^{ - 2}\left[ n \right]} {{\mathbf{\tilde h}}_{ru}}\left[ n \right],
\end{equation}
where ${{\mathbf{\tilde h}}_{ru}}\left[ n \right]$ denotes the phase of the channel, which is expressed as
\begin{equation}\small
	{{\mathbf{\tilde h}}_{ru}}\left[ n \right]={{{e}}^{{{ - j2\pi {d_{ru}}\left[ n \right]} \mathord{\left/
	{\vphantom {{ - j2\pi {d_{ru}}\left[ n \right]} \lambda }} \right.
	\kern-\nulldelimiterspace} \lambda }}}{u_x}\left[ n \right] \otimes {u_z}\left[ n \right],
\end{equation}
where
${u_x^T}\left[ n \right] \!=\! \left[ {1\!,\!{e^{ - j{{{2\pi d{\phi _{ru,x}}\left[ n \right]} \mathord{\left/{\vphantom {{2\pi d{\phi _{ru,x}}\left[ n \right]} \lambda }} \right.
	\kern-\nulldelimiterspace} \lambda }}}},\!...\!,{e^{ - j\left( {{L_x} - 1} \right){{{2\pi d{\phi _{ru,x}}\left[ n \right]} \mathord{\left/{\vphantom {{2\pi d{\phi _{ru,x}}\left[ n \right]} \lambda }} \right.
	\kern-\nulldelimiterspace} \lambda }}}}} \right],
$ $
 {u_z^T}\left[ n \right] \!=\! \left[ {1,{e^{ - j{{{2\pi d{\phi _{ru,z}}\left[ n \right]} \mathord{\left/{\vphantom {{2\pi d{\phi _{ru,z}}\left[ n \right]} \lambda }} \right.
	\kern-\nulldelimiterspace} \lambda }}}},...,{e^{ - j\left( {{L_z} - 1} \right){{{2\pi d{\phi _{ru,z}}\left[ n \right]} \mathord{\left/{\vphantom {{2\pi d{\phi _{ru,z}}\left[ n \right]} \lambda }} \right.
	\kern-\nulldelimiterspace} \lambda }}}}} \right] \!,\!
$ $
	{{\phi _{ru,x}}}\left[ n \right] = \sin \vartheta _{ru}^{\left( v \right)}\left[ n \right]\cos \vartheta _{ru}^{\left( h \right)}\left[ n \right] = {{\left( {x\left[ n \right] - {x_r}} \right)} \mathord{\left/
			{\vphantom {{\left( {x\left[ n \right] - {x_r}} \right)} {{d_{ru}}\left[ n \right]}}} \right.
			\kern-\nulldelimiterspace} {{d_{ru}}\left[ n \right]}},
$ $
	{\phi _{ru,z}}\left[ n \right] = \sin \vartheta _{ru}^{\left( v \right)}\left[ n \right]\sin \vartheta _{ru}^{\left( h \right)}\left[ n \right] = {{\left( {{H_u} - {z_r}} \right)} \mathord{\left/
			{\vphantom {{\left( {{H_u} - {z_r}} \right)} {{d_{ru}}\left[ n \right]}}} \right.
			\kern-\nulldelimiterspace} {{d_{ru}}\left[ n \right]}},
$
$\vartheta _{ru}^{\left( v \right)}$ and $\vartheta _{ru}^{\left( h \right)}$ represent the vertical and horizontal angle of departure (AoD) at the IRS, respectively.
${d_{ru}}\left[ n \right] = \left\| {{\mathbf{q}}\left[ n \right] - {{\mathbf{q}}_r}} \right\|$ is the distance between the IRS and the UAV. $d$ is the antenna distance.
The same model is adopted for GN-IRS channel and jammer-IRS channel. Then the channel gain is expressed as
\begin{equation}\label{5}
	{g_0}\left[ n \right]{\text{ = }}{\left| {{h_{gu}}\left[ n \right] + {\mathbf{h}}_{gr}^H\left[ n \right]{\mathbf{\Gamma }}\left[ n \right]{{\mathbf{h}}_{ru}}\left[ n \right]} \right|^2},
\end{equation}
\begin{equation}\label{6}
	{g_m}\left[ n \right]{\text{ = }}{\left| {{h_{mu}}\left[ n \right] + {\mathbf{h}}_{mr}^H\left[ n \right]{\mathbf{\Gamma }}\left[ n \right]{{\mathbf{h}}_{ru}}\left[ n \right]} \right|^2}.
\end{equation}
Then, the achievable average rate is given by 
\begin{equation}\label{R0}\small
	R = \frac{1}{N}\sum\limits_{n \in \cal{N}} {{{\log }_2}\left( {1 + \frac{{p[n]{g_0}\left[ n \right]}}{{{p_m}{g_m}\left[ n \right] + {\sigma ^2}}}} \right)} ,
\end{equation}
where ${p_m}$ denote the transmit power of the jammer, ${\sigma ^2}$ denotes the power of additive white Gaussian noise (AWGN) at the receiver. Thus, the problem is formulated as 
\begin{equation*}\small
	\begin{split}{\left( {{\rm{P0}}} \right)}
	:{\rm{ }}&\mathop {\max}\limits_{\bf{P},\bf{Q},{\mathbf{\Gamma }} } {R}\\
	{\rm{      }}{\rm s.t}.& ~ {\theta _i}[n] \in [0,2\pi ),i \in \{ 1,...,L\} ,\forall n,\\
	&~\left( {\rm{1}} \right)- \left( {\rm{6}} \right).		
    \end{split}
\end{equation*}
It is challenging to solve (P0) optimally due to the non-convex objective function and coupled optimization variables. However, it can be effectively solved by dividing the problem into three sub-problems with consideration for the worst case. Thus, we exploit an alternating optimization (AO) to ensure that the sub-problems are convex in each iteration, while fixing the other two in each iteration until convergence is achieved.
			
\section{The Proposed Alternating Algorithm}
	\label{The Proposed Alternating Algorithm}
	\subsection{Sub-Problem 1: Optimizing {\bf P} for Given {\bf Q} and $\bf \Gamma$ }
			
For given the UAV trajectory ${\bf{Q}}$ and the phase shift ${\mathbf{\Gamma }} $, we consider the worst situation to deal with the jammer's uncertain location. When the jammer source is located closest to the UAV, we can obtain the lower bound of the average rate. $(\rm{P0})$ can be rewritten as 
\begin{equation*}
	\begin{split}{\left( {{\rm{P1}}} \right)}
	:{\rm{ }}&\mathop {\max }\limits_{\bf{P}} \frac{1}{N}\sum\limits_{n \in {\cal{N}}} {{{\log }_2}\left( {1 + \frac{{p[n]{{g_0}\left[ n \right]}}}{{p_m}{{{{\tilde g}_m}}}\left[ n \right] + {\sigma ^2}}} \right)}\\
	&{\rm{s}}{\rm{.t}}{\rm{.}}~\left( {\rm{5}} \right), \left( {\rm{6}} \right).
	\end{split}
\end{equation*}		
This is a standard convex optimization problem that can be efficiently solved by some existing algorithms, such as CVX's interior point method.
					
\subsection{Sub-Problem 2: Optimizing $\bf \Gamma$ for Given {\bf Q} and {\bf P}}
					
For given trajectory ${\bf{Q}}$ and transmit power ${\bf{P}}$, $(\rm{P0})$ can be transformed as 
\begin{equation*}
	\begin{gathered}
	\left( {{\rm{P}}2} \right):\mathop {\max }\limits_{\bf{\Gamma} } \frac{1}{N}\sum\limits_{n \in \cal{N}} {{{\log }_2}\left( {1 + \frac{{p[n]{g_0}\left[ n \right]}}{{{p_m}{g_m}\left[ n \right] + {\sigma ^2}}}} \right)}  \hfill \\
	~~~~~~~~{\text{s}}{\text{.t}}{\text{.}}~{\theta _i}[n] \in [0,2\pi ),i \in \{ 1,...,L\} ,\forall n, \hfill \\ 
	\end{gathered} 
\end{equation*}
it is hard to obtain the optimal solution due to the unit modulus constraints and the imperfect CSI caused by jammer's uncertain location. Thus, we consider a practical scheme to estimate the ${{\mathbf{h}}_{mr}}$ and ${{\mathbf{h}}_{mu}}$, which belong to a given range, i.e.
\begin{equation*}\small
	\begin{gathered}
	{{\mathbf{\Psi }}_1} = \left\{ {{{\mathbf{h}}_{mr}}\left| {\left| {{\beta _{mr}}} \right| \in \left[ {\beta _{mr}^{\min },\beta _{mr}^{\max }} \right]} \right.,} \right. \hfill \\
	\left. {\phi _{mr,i} \in \left[ {{{\left( {\phi _{mr,i}} \right)}^{\min }},{{\left( {\phi _{mr,i}^{\left( i \right)}} \right)}^{\max }}} \right],i \in \left\{ {x,z} \right\}} \right\}, \hfill \\ 
\end{gathered}
\end{equation*}
\begin{equation*}\small
{{\mathbf{\Psi }}_2} = \left\{ {{{\mathbf{h}}_{mu}}\left| {\left| {{\beta _{mu}}} \right| \in \left[ {\beta _{mu}^{\min },\beta _{mu}^{\max }} \right]} \right.} \right\},
\end{equation*}
where ${\beta _{mu}}$, ${\beta _{mr}}$ denote the amplitude of path loss, superscripts min and max respectively denote the lower and upper bound. We denote ${\mathbf {\hat v}}^H[n] = [{e^{j{\theta _1}[n]}},{e^{j{\theta _2}[n]}},...,{e^{j{\theta _L}[n]}}]$, ${\mathbf v}^{H}\left[ n \right] = {e^{j\tau  }} \left[{\mathbf {\hat v}^H}\left[ n\right],1 \right]$, where $\tau$ is an arbitrary phase rotation. As such, (\ref{5}) and (\ref{6}) can be rewritten as 
\begin{equation*}\small
	{g_0}\left[ n \right] = {\left| {{{\mathbf{v}}^H}\left[ n \right]{\text{diag}}\left\{ {{{\mathbf{h}}_0}\left[ n \right]} \right\}{{\mathbf{h}}_g}\left[ n \right]} \right|^2},
\end{equation*}
\begin{equation*}\small
	{g_m}\left[ n \right] = {\left| {{{\mathbf{v}}^H}\left[ n \right]{\text{diag}}\left\{ {{{\mathbf{h}}_0}\left[ n \right]} \right\}{{\mathbf{h}}_m}\left[ n \right]} \right|^2},
\end{equation*}
where ${{\mathbf{h}}_i}\left[ n \right] = {\left[ {{\mathbf{h}}_{ir}^H\left[ n \right],{h_{iu}}\left[ n \right]} \right]^H},i \in \left\{ {m,g} \right\} $, ${{\mathbf{h}}_0}\left[ n \right] = {\left[ {{\mathbf{h}}_{ru}^H\left[ n \right],1} \right]^H} $. (P0) can be equivalently written as
\begin{equation*}
	\begin{gathered}
	\left( {{\rm{P}}2.1} \right):\mathop {\max }\limits_{\mathbf{v}} \mathop {\min }\limits_{{{\mathbf{w}}_m}} \frac{{{{\mathbf{v}}^H}\left[ n \right]{{\mathbf{w}}_g}\left[ n \right]{\mathbf{w}}_g^H\left[ n \right]{\mathbf{v}}\left[ n \right]}}{{{{\mathbf{v}}^H}\left[ n \right]{{\mathbf{w}}_m}\left[ n \right]{\mathbf{w}}_m^H\left[ n \right]{\mathbf{v}}\left[ n \right] + {\sigma ^2}}}\\
	~~~~~~~~{\text{s}}{\text{.t}}{\text{.}}~{\theta _i}[n] \in [0,2\pi ),i \in \{ 1,...,L\} ,\forall n, \hfill \\ 
    \end{gathered} 
\end{equation*}
where ${{\mathbf{w}}_m}\left[ n \right] = \sqrt {{p_m}} {\text{diag}}\left\{ {{{\mathbf{h}}_0}\left[ n \right]} \right\}{{\mathbf{h}}_m}\left[ n \right]$, ${{\mathbf{w}}_g}\left[ n \right] = \sqrt {p\left[ n \right]} {\text{diag}}\left\{ {{{\mathbf{h}}_0}\left[ n \right]} \right\}{{\mathbf{h}}_g}\left[ n \right]$.
					
To overcome the difficulty of the imperfect CSI, we consider the method which constructs a convex hull based on weighted sum of $T$ discrete samples i.e. \cite{24}
\begin{equation*}\small
\Omega {\text{ = }}\left\{ {\sum\limits_{t = 1}^T {{\alpha _t}{\mathbf{h}}_t^H{{\mathbf{h}}_t}} \left| {\sum\limits_{t = 1}^T {{\alpha _t} = 1,} } \right.{\alpha _t} \geqslant 0} \right\},
\end{equation*}
where ${{\mathbf{h}}_t} = {\left( {{\text{diag}}\left\{ {{{\mathbf{h}}_0}} \right\}{{\mathbf{h}}_m}} \right)_t}$ and ${\alpha _t}$ denotes the weighted coefficient of the $t$-th discrete sample. (P2.1) is written as 
\begin{equation*}
	\begin{gathered}
	\left( {{\rm{P}}2.2} \right):\mathop {\max }\limits_{\mathbf{v}} \mathop {\min }\limits_\Omega  \frac{{{{\mathbf{v}}^H}\left[ n \right]{{\mathbf{w}}_g}\left[ n \right]{\mathbf{w}}_g^H\left[ n \right]{\mathbf{v}}\left[ n \right]}}{{\sum\limits_{t = 1}^T {{\alpha _t}\left[ n \right]} {{\mathbf{v}}^H}\left[ n \right]{{\mathbf{h}}_t}\left[ n \right]{\mathbf{h}}_t^H\left[ n \right]{\mathbf{v}}\left[ n \right] + {\sigma ^2}}}\\
	~~~~~~~~{\text{s}}{\text{.t}}{\text{.}}~{\theta _i}[n] \in [0,2\pi ),i \in \{ 1,...,L\} ,\forall n. \hfill \\ 
	\end{gathered} 
\end{equation*}				
(P2.2) is still an intractable problem. However, it's observed that the worst case of the objection can be efficiently solved when ${\mathbf{v}}$ is fixed. When  ${\mathbf{v}}$ is fixed, (P2.2) can be rewritten as				
\begin{equation*}\small
	\mathop {\min }\limits_\Omega  \frac{{{{\mathbf{v}}^H}\left[ n \right]{{\mathbf{w}}_g}\left[ n \right]{\mathbf{w}}_g^H\left[ n \right]{\mathbf{v}}\left[ n \right]}}{{\sum\limits_{t = 1}^T {{\alpha _t}} {{\mathbf{v}}^H}\left[ n \right]{{\mathbf{h}}_t}\left[ n \right]{\mathbf{h}}_t^H\left[ n \right]{\mathbf{v}}\left[ n \right] + {\sigma ^2}}}.
\end{equation*}
Let ${{\mathbf{\bar w}}_t}\left[ n \right] = {{\mathbf{v}}^H}\left[ n \right]{{\mathbf{h}}_t}\left[ n \right]$, the optimal weighted coefficient is obtained by maximizing $\sum\limits_{t = 1}^T {{\alpha _t}} {\mathbf{\bar w}}_t^H\left[ n \right]{{\mathbf{\bar w}}_t}\left[ n \right]$. By using the Cauchy-Schwarz's inequality, we have 
\begin{equation}\label{11}\small
	\begin{gathered}
	{\left( {\sum\limits_{t = 1}^T {{\alpha _t}\left[ n \right]} {\mathbf{\bar w}}_t^H\left[ n \right]{{{\mathbf{\bar w}}}_t}\left[ n \right]} \right)^2} \!\leqslant\! \left( {\sum\limits_{t = 1}^T {\alpha _t^2\left[ n \right]} } \right)\sum\limits_{t = 1}^T {{{\left( {{\mathbf{\bar w}}_t^H\left[ n \right]{{{\mathbf{\bar w}}}_t}\left[ n \right]} \right)}^2}}.  \hfill \\ 
	\end{gathered} 
\end{equation}			
The inequality (\ref{11})  holds equal only when  ${{{\alpha _1}\left[ n \right]} \mathord{\left/
{\vphantom {{{\alpha _1}\left[ n \right]} {{\mathbf{\bar w}}_1^H\left[ n \right]{{{\mathbf{\bar w}}}_1}\left[ n \right]}}} \right.
\kern-\nulldelimiterspace} {{\mathbf{\bar w}}_1^H\left[ n \right]{{{\mathbf{\bar w}}}_1}\left[ n \right]}} = {{{\alpha _2}\left[ n \right]} \mathord{\left/
{\vphantom {{{\alpha _2}\left[ n \right]} {{\mathbf{\bar w}}_2^H\left[ n \right]{{{\mathbf{\bar w}}}_2}\left[ n \right]}}} \right.
\kern-\nulldelimiterspace} {{\mathbf{\bar w}}_2^H\left[ n \right]{{{\mathbf{\bar w}}}_2}\left[ n \right]}} = ... = {{{\alpha _T}\left[ n \right]} \mathord{\left/
{\vphantom {{{\alpha _T}\left[ n \right]} {{\mathbf{\bar w}}_T^H\left[ n \right]{{{\mathbf{\bar w}}}_T}\left[ n \right]}}} \right.
\kern-\nulldelimiterspace} {{\mathbf{\bar w}}_T^H\left[ n \right]{{{\mathbf{\bar w}}}_T}\left[ n \right]}}$ is satisfied. Since we have $\sum\limits_{t = 1}^T {{\alpha _t} = 1}$, ${\alpha _t}$ can be obtained as 					
\begin{equation}\label{12}\small
	{\alpha _t}\left[ n \right] = {\mathbf{\bar w}}_t^H\left[ n \right]{{\mathbf{\bar w}}_t}\left[ n \right]{\left( {\sum\limits_{t = 1}^T {{\mathbf{\bar w}}_t^H\left[ n \right]{{{\mathbf{\bar w}}}_t}\left[ n \right]} } \right)^{ - 1}}.
\end{equation}
Based on the given ${\alpha _t}$, (P2.2) can be transformed as
\begin{equation*}\small
	\begin{aligned}
	\left( {{\rm{P}}2.3} \right):&\mathop {\max }\limits_{\mathbf{v}} \frac{{{{\mathbf{v}}^H}\left[ n \right]{{\mathbf{w}}_g}\left[ n \right]{\mathbf{w}}_g^H\left[ n \right]{\mathbf{v}}\left[ n \right]}}{{\sum\limits_{t = 1}^T {{\alpha _t}\left[ n \right]} {{\mathbf{v}}^H}\left[ n \right]{{\mathbf{h}}_t}\left[ n \right]{\mathbf{h}}_t^H\left[ n \right]{\mathbf{v}}\left[ n \right] + {\sigma ^2}}} \hfill \\
	&{\text{s}}{\text{.t}}{\text{.}}~{\theta _i}[n] \in [0,2\pi ),i \in \{ 1,...,L\} ,\forall n. \hfill \\ 
\end{aligned}
\end{equation*}
Let ${\mathbf{V}}\left[ n \right] = {\mathbf{v}}\left[ n \right]{{\mathbf{v}}^H}\left[ n \right]$, ${\mathbf{A}}\left[ n \right] = {{\mathbf{w}}_g}\left[ n \right]{\mathbf{w}}_g^H\left[ n \right]$ and ${\mathbf{B}}\left[ n \right] = \sum\limits_{t = 1}^T {{\alpha _t}\left[ n \right]} {{\mathbf{\bar w}}_t}\left[ n \right]{\mathbf{\bar w}}_t^H\left[ n \right]$, then (P2.3) can be reformulated as
\begin{equation*}
\begin{aligned}
	\left( {{\rm{P}}2.4} \right):\mathop {\max }\limits_{\mathbf{V}}& \frac{{{\text{tr}}\left( {{\mathbf{A}}\left[ n \right]{\mathbf{V}}\left[ n \right]} \right)}}{{{\text{tr}}\left( {{\mathbf{B}}\left[ n \right]{\mathbf{V}}\left[ n \right]} \right) + {\sigma ^2}}} \hfill \\ 
	{\text{s}}{\text{.t}}{\text{.}}~&{{\mathbf{V}}_{l,l}}\left[ n \right] = 1,l = 1,2...,L + 1,\forall n, \hfill \\ 
	&{{\mathbf{V}}}\left[ n \right]\underset{\raise0.3em\hbox{$\smash{\scriptscriptstyle-}$}}{ \succ } 0,{\text{rank}}\left( {{\mathbf{V}}\left[ n \right]} \right) = 1,\forall n,
\end{aligned}
\end{equation*}
where ${{\mathbf{V}}_{l,l}}\left[ n \right]$ denotes the $(l,l)$-th element of ${\mathbf{V}}\left[ n \right]$. However, note that ${\text{rank}}\left( {{\mathbf{V}}\left[ n \right]} \right) = 1$ is non-convex, we consider applying SDR. By defining  $k\left[ n \right]{\text{ = }}{1 \mathord{\left/
{\vphantom {1 {\left( {{\text{tr}}\left( {{{\mathbf{0}}_2}{\mathbf{V}}\left[ n \right]} \right) + \sigma _M^2} \right)}}} \right.
\kern-\nulldelimiterspace} {\left( {{\text{tr}}\left( {{{\mathbf{0}}_2}{\mathbf{V}}\left[ n \right]} \right) + \sigma _M^2} \right)}}$ and ${\mathbf{\tilde V}}\left[ n \right] = k\left[ n \right]{\mathbf{V}}\left[ n \right]$, we can rewrite (P2.4) as a convex semidefinite programming (SDP) problem as follow
\begin{equation*}
	\begin{aligned}
	\left( {{\rm{P}}2.5} \right):\mathop {\max }\limits_{{\mathbf{\tilde V}},{\mathbf{k}}}& \quad {\text{tr}}\left( {{\mathbf{A}}\left[ n \right]{\mathbf{\tilde V}}\left[ n \right]} \right) \hfill \\ 
	{\text{s}}{\text{.t}}{\text{.}}~&{{\mathbf{V}}_{l,l}}\left[ n \right] = k\left[ n \right],l = 1,2...,L + 1,\forall n, \hfill \\ 
	&{{\mathbf{\tilde V}}_{l,l}}\left[ n \right]\underset{\raise0.3em\hbox{$\smash{\scriptscriptstyle-}$}}{ \succ } 0,k \left[ n \right] \geqslant 0,\forall n, \hfill \\ 
	&{\text{tr}}\left( {{\mathbf{B}}\left[ n \right]{\mathbf{\tilde V}}\left[ n \right]} \right) + k\left[ n \right]{\sigma ^2} = 1,\forall n,
\end{aligned}
\end{equation*}
(P2.5) can be efficiently solved by using convex optimization toolbox such as CVX. However, the constraint ${\text{rank}}\left( {{\mathbf{V}}\left[ n \right]} \right) = 1$ may be not always guaranteed. Specifically, if the obtained ${{\mathbf{V}}\left[ n \right]}$ is of rank-1, we can obtain the ${{\mathbf{v}}\left[ n \right]}$ by applying
eigenvalue decomposition, thus the obtained ${{\mathbf{v}}\left[ n \right]}$ is the optimal solution to (P2.3). Otherwise, Gaussian randomization is
needed for recovering ${{\mathbf{v}}\left[ n \right]}$ approximately. The details can be found in \cite{5} and thus are omitted here. Therefore, the coefficients are obtained as 
\begin{equation*}
	{\hat v_i}\left[ n \right] = {e^{j\angle \left( {\frac{{{v_i}\left[ n \right]}}{{{v_{N + 1}}\left[ n \right]}}} \right)}},n = 1,...,L.
\end{equation*}
The proposed algorithm to solve the (P2) is summarized as {\bf{Algorithm  1}}, where ${R_{irs}}$ is the objective function of (P2).
					
\begin{algorithm}[t]\small
	\renewcommand{\algorithmicrequire}{\textbf{Input:}}
	\renewcommand{\algorithmicensure}{\textbf{Output:}}
	\caption{Algorithm for (P2) in each time slot}\label{Algorithm1}
	\begin{algorithmic}[1]
	\STATE\textbf{Input:}
${\mathbf{P}}$, ${\mathbf{Q}}$, $T$, ${\mu _1}$. 
	\STATE\textbf{Onput:} ${{\mathbf{v}}\left[ n \right]}$.
	\STATE Initialize ${{\mathbf{v}}\left[ n \right]}=I_{L+1}, i=0$, $R_{irs}^{\left( 0 \right)}=0$.
	\STATE Let ${\varepsilon _{m,t}} = \left[ {\vartriangle {x_{m,t}},\vartriangle {y_{m,t}},\vartriangle {z_{m,t}}} \right]$, ${\beta _{mr}} = \beta _{mr}^{\min }$, and ${\beta _{mu}} = \beta _{mu}^{\min }$, compute ${\phi _{x,t}}\left[ n \right]$, ${\phi _{z,t}}\left[ n \right]$ and obtain ${{\mathbf{\tilde h}}_{mr,t}}$ and ${\tilde h_{mu,t}}$. Then construct ${{{\mathbf{h}}_t}}\left[ n \right]$, where $t = 1,2,...T$.
	\STATE\textbf{~repeat:}
	\STATE~ i:=i+1.
	\STATE~ Compute $\alpha _t^{\left( i \right)}$ based on (\ref{12}).
	\STATE~ Compute ${{\mathbf{v}}\left[ n \right]}$ for given $\alpha _t^{\left( i \right)}\left[ n \right]$ by solving (P2.5).
	\STATE{\textbf{~until} $\left| {R_{irs}^{\left( i \right)}\left[ n \right] - R_{irs}^{\left( {i - 1} \right)}} \left[ n \right]\right| \leqslant {\mu _1}$.  }
	\end{algorithmic}
\end{algorithm}			
\subsection{Sub-Problem 3: Optimizing {\bf Q} for Given {\bf P} and ${\bf \Gamma}$}
For given ${\bf{P}}$ and $\bf{\Gamma} $, (P0) can be rewritten as 
\begin{equation*}
	\begin{split}{\left( {{\rm{P3}}} \right)}
		:{\rm{ }}&\mathop {\max}\limits_{\bf{Q} } {R}\\
		{\rm{      }}{\rm s.t}.&~\left( {\rm{1}} \right)-\left( {\rm{4}} \right),		
    \end{split}
\end{equation*}
(P3) is difficult to solve due to the uncertain location of the jammer and its non-convex objective function. To overcome the imperfect location of the jammer, we consider introducing the slack variables ${\mathbf{D}}{\text{ = }}\left\{ {d[n],\forall n} \right\}$, which are used to approximately denote the square of distance from the jammer to the UAV. For the jammer-IRS channel, we just consider the worst case that the jammer is farthest to the IRS. As such, we can conveniently to transform the problem into a convex form, which is given by
\begin{equation}\label{15}
	d[n] \leqslant {\left\| {{\mathbf{q}}[n] - {{\mathbf{q}}_m}} \right\|^2},d[n] \geqslant 0,\forall n,\\
\end{equation}
where ${q_m}$ contains infinite number of variables due to the uncertain location information of the jammer. By resorting to $\mathcal{S} - procedure$, the constraints of imperfect location information can be rewritten as
\begin{equation*}\small
\Delta x_m^2 + \Delta y_m^2 + \Delta z_m^2 - D_m^2 \leqslant 0,
\end{equation*}
\begin{equation*}
	\begin{aligned}
	- {\left( {{{\bar x}_m} + \Delta x_m - x[n]} \right)^2} - {\left( {{\bar y}_m} + \Delta y_m - y[n] \right)^2} - \\~~{\left( {{{\bar z}_m} + \Delta z_m - z[n]} \right)^2} + d[n] \leqslant 0,
	\end{aligned}
\end{equation*}
and there exists $\left( {\Delta {{\hat x}_m},\Delta {{\hat y}_m},\Delta {{\hat z}_m}} \right) = \left( {0,0,0} \right)$ satisfying constraint. Therefore, according to \textbf{ \emph{Lemma}} \emph{1}, we introduce ${\mathbf{\delta }} \triangleq \left\{ {\delta \left[ n \right] \geqslant 0,\forall n} \right\}$, the equality constraint is given by 
\begin{equation}\label{16}
	\begin{gathered}
	\Theta (x[n],y[n],d[n],\delta [n])= \hfill \\
	\left[ {\begin{array}{*{20}{c}}
	{\delta \left[ n \right] + 1}&0&0&{{{\bar x}_m} - x[n]} \\ 
	0&{\delta \left[ n \right] + 1}&0&{{{\bar y}_m} - y[n]} \\ 
	0&0&{\delta \left[ n \right] + 1}&{{{\bar z}_m} - H_u} \\ 
	{{{\bar x}_m} - x[n]}&{{{\bar y}_m} - y[n]}&{{{\bar z}_m} - H_u}&{E[n]} 
	\end{array}} \right], \hfill \\ 
	\end{gathered} 
\end{equation}
where
$E\left[ n \right] =  - D_m^2\delta [n] + {x^2}[n] - 2{\bar x_m}x[n] + \bar x_m^2 + \\{y^2}[n] - 2{\bar y_m}y[n] + \bar y_m^2 + {H_u^2}[n] - 2{\bar z_m}H_u[n] + \bar z_m^2 - d[n],\forall n$. To make the problem solvable, we use the first-order Taylor expansion of the convex function to obtain the lower bound. At the given feasible points ${{\mathbf{x}}_0} = \left\{ {{x_0}\left[ n \right],\forall n} \right\}$ and ${{\mathbf{y}}_0} = \left\{ {{y_0}\left[ n \right],\forall n} \right\}$, $E[n]$ can be transformed as 				
\begin{equation*}
\begin{gathered}
	\tilde E\left[ n \right]=  \!-\! D_m^2\delta \left[ n \right] - x_0^2\left[ n \right] \!+\! 2{x_0}\left[ n \right]x\left[ n \right] \!-\! 2{{\bar x}_m}x[n]
	+ \bar x_m^2 - y_0^2[n] \\  + 2{y_0}\left[ n \right]y\left[ n \right] - 2{{\bar y}_m}y[n] + \bar y_m^2+H_u^2-2{{\bar z}_m}z[n] + \bar z_m^2 - d[n]. \\ 
	\end{gathered} 
\end{equation*}
Then we substitute the $\tilde E\left[ n \right]$ into (\ref{16}) and obtain the convex formation
\begin{equation}
	\tilde \Theta (x[n],y[n],d[n],\delta [n])\underset{\raise0.3em\hbox{$\smash{\scriptscriptstyle-}$}}{ \succ } 0,
\end{equation}
which is a semidefinite programming problem that can be solved optimally by CVX. Next, to deal with the non-convex objective function of (P3), we observe that ${{{\mathbf{\tilde h}}}_{ru}}\left[ n \right]$, ${{{\mathbf{\tilde h}}}_{gu}}\left[ n \right]$ and ${{{\mathbf{\tilde h}}}_{mu}}\left[ n \right]$ are relative to the trajectory variables. However,it is observed that those are more complex to handle due to its non-linear formation. To overcome such problem, we use ($j-1$)th iteration to obtain the approximate ${{{\mathbf{\tilde h}}}_{ru}}\left[ n \right]$,${{{\mathbf{\tilde h}}}_{gu}}\left[ n \right]$ and ${{{\mathbf{\tilde h}}}_{mu}}\left[ n \right]$ \cite{20}. Based on the consideration above, we consider rewriting the objective function. By denoting
\begin{equation*}
\begin{gathered}
	{{\bf{H}}_g}\left[ n \right] = \left[ {\sqrt \rho  \tilde h_{gu}^{\left(i-1\right)}\left[ n \right],\sqrt \rho  {\mathbf{h}}_{gr}^H\left[ n \right]{\mathbf{\Gamma }}\left[ n \right]{{{\mathbf{\tilde h}}}_{ru}^{\left(i-1\right)}}\left[ n \right]} \right] \hfill \\
	{{\bf{H}}_m}\left[ n \right] = \left[ {\sqrt \rho  \tilde h_{mu}^{\left(i-1\right)}\left[ n \right],\sqrt \rho  {\mathbf{ h}}_{mr}^H\left[ n \right]{\mathbf{\Gamma }}\left[ n \right]{{{\mathbf{\tilde h}}}_{ru}^{\left(i-1\right)}}\left[ n \right]} \right], \hfill \\ 
\end{gathered} 
\end{equation*}
where  
${\tilde h_{gu}}[n] = {e^{ - j\frac{{2\pi {d_{gu}}\left[ n \right]}}{\lambda }}}$, (\ref{5}), (\ref{6}) can be transformed as				
\begin{equation*}
	{g_0}\left[ n \right] = {\mathbf{d}}_g^T[n]{\mathbf{H}}_g^H[n]{{\mathbf{H}}_g}[n]{{\mathbf{d}}_g}[n],
\end{equation*}
\begin{equation*}
	{g_m}\left[ n \right] = {\mathbf{d}}_m^T[n]{\mathbf{H}}_m^H[n]{{\mathbf{H}}_m}[n]{{\mathbf{d}}_m}[n],
\end{equation*}
where ${{\mathbf{d}}_g}[n] = {\left[ {{ {{d_{gu}^{-1}}[n]} } ,{{{d_{ru}^{ - 1}}[n]} } } \right]^T}$, ${{\mathbf{d}}_m}[n] = {\text{ }}{\left[ {\sqrt {{ {d^{ - 1}[n]} }} , {{{d_{ru}^{ - 1}}[n]}} } \right]^T}$. 
Moreover, the other problem is the non-convex objective function which is composed of the coupled variables. To tackle it, we consider introducing the slack variables and leveraging the SCA method. By introducing the slack variables  ${\mathbf{S}} = \{ S[n],\forall n\}$ and ${\mathbf{G}} = \{ G[n],\forall n\}$, an optimization problem equivalent to (P3) is obtained as follows
\begin{equation*}\small
\begin{gathered}
	\left( {{\rm{P}}3.1} \right):\mathop {\max }\limits_{{\mathbf{Q}},{\mathbf{S}},{\mathbf{G}}} \frac{1}{N}\sum\limits_{n \in {\text{N}}} {{{\log }_2}\left( {1 + \frac{{{S^{ - 1}}[n]}}{{G[n]}}} \right)}  \hfill \\
	~~~~~~~{\text{s.t.}}~{\rm C}1:p[n]{\mathbf{d}}_g^T[n]{\mathbf{H}}_g^H[n]{{\mathbf{H}}_g}[n]{{\mathbf{d}}_G}[n] \geqslant {S^{ - 1}}[n],\forall n, \hfill \\
    ~~~~~~~\quad ~~{\rm C}2:{p_m}{\mathbf{d}}_m^T[n]{\mathbf{H}}_m^H[n]{{\mathbf{H}}_m}[n]{{\mathbf{d}}_m}[n] + {\sigma ^2} \leqslant G[n],\forall n. \hfill \\ 
    ~~~~~~~~~~~\left( 3 \right)-\left( 4 \right), \left( 18 \right). \hfill \\
\end{gathered} 
\end{equation*}
(P3) and (P3.1) share the same optimal solution when the constraints hold with equalities \cite{3}. And the objective function in (P3) has a lower bound at $\left( {{S_0}\left[ n \right],{G_0}\left[ n \right]} \right)$ by 
\begin{equation*}\small
\begin{gathered}
	\tilde R\left( {S[n],G[n]} \right) = {\text{l}}o{g_2}\left( {1 + {1 \mathord{\left/
				{\vphantom {1 {\left( {{S_0}[n]{G_0}[n]} \right)}}} \right.
				\kern-\nulldelimiterspace} {\left( {{S_0}[n]{G_0}[n]} \right)}}} \right) \hfill \\ + {\zeta _1}\left[ n \right]\left( {S\left[ n \right] - {S_0}\left[ n \right]} \right) + {\zeta _2}\left[ n \right]\left( {G\left[ n \right] - {G_0}\left[ n \right]} \right) \hfill, \\ 
\end{gathered}
\end{equation*}
where ${\zeta _1}[n] =  - {\log _2}\left( {\frac{e}{{{S_0}[n] + {{({S_0}[n])}^2}{G_0}[n]}}} \right)$, ${\zeta _2}[n] =  - {\log _2}\left( {\frac{e}{{{G_0}[n] + {{({G_0}[n])}^2}{S_0}[n]}}} \right)$. Thus (P3.1) is lower bounded by 
\begin{equation*}
	\begin{gathered}
		\left( {{\rm{P}}3.2} \right):\mathop {\max }\limits_{{\mathbf{Q}},{\mathbf{S}},{\mathbf{G}}} \frac{1}{N}\sum\limits_{n \in {\cal{N}}} {{\tilde{R}\left( {S[n],G[n]} \right)}} \hfill \\
		~~~~{\text{s}}{\text{.t}}{\text{.}}
		~{\rm C}1, {\rm C}2, \left( 3 \right)- \left( 4 \right), \left( 18 \right). \hfill \\
	\end{gathered} 
\end{equation*}
However, the constraints ${\rm C}1$ and ${\rm C}2$ are non-convex because of the coupled variables. To deal with the non-convexity, we introduce the slack variables ${{\bf{\xi }}_1} = \left\{ {{\xi _1}\left[ n \right],\forall n} \right\}$, ${{\mathbf{\xi }}_2} = \left\{ {{\xi _2}\left[ n \right],\forall n} \right\}$, ${{\mathbf{\xi }}_3} = \left\{ {{\xi _3}\left[ n \right],\forall n} \right\}$, ${{\mathbf{\xi }}_4} = \left\{ {{\xi _4}\left[ n \right],\forall n} \right\}$. The problem is transformed into 
\begin{equation*}
    \begin{split}{\left( {{\rm{P3.3}}} \right)}
	&:\mathop {\max }\limits_{~~~~\bf{Q},S,G, \hfill\atop
	~	 {\xi _1}, {\xi _2}, {\xi _3}, {\xi _4}\hfill} \frac{1}{N}\sum\limits_{n \in {\cal{N}}} {{\tilde{R}\left( {S[n],G[n]} \right)}} \\
	&{\rm{s}}{\rm{.t}}{\rm{.}}~{\rm C}1.1:p[n]\tilde {\bf{d}}_{g}^T[n]{\bf{H}}_{g}^H[n]{{\bf{H}}_{g}}[n]{{\tilde {\bf{d}}}_{g}}[n] \ge {S^{ - 1}}[n],\\
	&~~~~~{\rm C}2.1:{p_m}\tilde {\bf{d}}_{m}^T[n]{\bf{H}}_{g}^H[n]{{\bf{H}}_{g}}[n]{{\tilde {\bf{d}}}_{m}}[n] + {\sigma ^2}\le G[n],\\
	&~~~~~ {d_{gu}^{ -1 }}[n]  \ge {\xi _1}[n],{d_{ru}^{ -1 }}[n]  \ge  {\xi _2}[n],\forall n,\\
	&~~~~~\sqrt {{d}^{ -1 }[n]}  \le  {\xi _3}[n],{d_{ru}^{ -1 }}[n]  \le {\xi _4}[n],\forall n,\\
	&~~~~~\left( 3 \right)-\left( 4 \right),\left( 18 \right),
	\end{split}
\end{equation*}
where ${{\mathbf{\tilde d}}_g} = {\left[ {{\xi _1}\left[ n \right],{\xi _2}\left[ n \right]} \right]^T}$, ${{\mathbf{\tilde d}}_m} = {\left[ {{\xi _3}\left[ n \right],{\xi _4}\left[ n \right]} \right]^T}$.	
To conveniently tackle the non-convex variables, we unfold as
\begin{equation}\label{18}
	\begin{gathered}
    {F_1}\left[ n \right] - {{\xi _1^{ - 2}\left[ n \right]}} \leqslant 0,{F_2}\left[ n \right] - {{\xi _2^{ - 2}\left[ n \right]} } \leqslant 0,\forall n, \hfill \\
	{\xi _3^{ - 2}\left[ n \right]}  - d\left[ n \right] \leqslant 0,{\xi _4^{ - 2}\left[ n \right]}  - {F_2}\left[ n \right] \leqslant 0,\forall n, \hfill \\ 
	\end{gathered}
\end{equation}
where ${F_1}\left[ n \right] = {\left( {x\left[ n \right] - {x_g}} \right)^2} + {\left( {y\left[ n \right] - {y_g}} \right)^2} + {H_u^2}$, ${F_2}\left[ n \right] = {\left( {x\left[ n \right] - {x_r}} \right)^2} + {\left( {y\left[ n \right] - {y_r}} \right)^2} + {\left( {H_u - {z_r}} \right)^2}$. However, there still exists several non-convex feasible regions in constraints (\ref{18}). The first-order Taylor expansions of ${ {\xi _1^{ - 2}\left[ n \right]}}$, ${{\xi _2^{ - 2}\left[ n \right]} }$, ${\mathbf{d}}_g^T[n]{\mathbf{H}}_g^H[n]{{\mathbf{H}}_g}[n]{{\mathbf{d}}_g}[n]$ at the feasible points ${{\mathbf{\xi }}_{{\mathbf{1}},{\mathbf{0}}}} = \left\{ {{\xi _{1,0}}\left[ n \right],\forall n} \right\}$, ${{\mathbf{\xi }}_{{\mathbf{2}},{\mathbf{0}}}} = \left\{ {{\xi _{2,0}}\left[ n \right],\forall n} \right\}$ and ${{\mathbf{\tilde d}}_{g,0}} = \left\{ {{{\tilde d}_{g,0}}[n],\forall n} \right\}$, which is given by
\begin{equation*}\small
	\begin{split}
		&\xi _1^{ - 2}\left[ n \right] \geqslant \xi _{1,0}^{ - 2}\left[ n \right] - 2\xi _{1,0}^{ - 2 - 1}\left[ n \right]\left( {{\xi _1}\left[ n \right] - {\xi _{1,0}}\left[ n \right]} \right)={ {{{\tilde \xi }_1}\left[ n \right]} },\hfill\\
		&\xi _2^{ - 2}\left[ n \right] \geqslant \xi _{2,0}^{ - 2}\left[ n \right] - 2\xi _{2,0}^{ - 2 - 1}\left[ n \right]\left( {{\xi _2}\left[ n \right] - {\xi _{2,0}}\left[ n \right]} \right)={ {{{\tilde \xi }_2}\left[ n \right]} },\hfill\\
		\end{split} 
\end{equation*} 
\begin{equation*}\small
	\begin{split}
		&{\mathbf{\tilde d}}_g^T[n]{\mathbf{H}}_g^H[n]{{\mathbf{H}}_g}[n]{{{\mathbf{\tilde d}}}_g}[n] \geqslant  - {\mathbf{\tilde d}}_{g,0}^T[n]{\mathbf{H}}_g^H[n]{{\mathbf{H}}_g}[n]{{{\mathbf{\tilde d}}}_{g,0}}[n] \hfill\\
		&+ 2\Re \left[ {{\mathbf{\tilde d}}_{g,0}^T[n]{\mathbf{H}}_g^H[n]{{\mathbf{H}}_{\text{g}}}[n]{{{\mathbf{\tilde d}}}_g}[n]} \right], \hfill \\ 
	\end{split} 
\end{equation*} 
and ${F_2}\left[ n \right]$ can be transformed as
\begin{equation*}
	\begin{gathered}
		{{\tilde F}_2}\left[ n \right] =  - x_0^2\left[ n \right] + 2{x_0}\left[ n \right]x\left[ n \right] + x_r^2 - y_0^2\left[ n \right] + \\ 2{y_0}\left[ n \right]y\left[ n \right]-y_r^2 \!+\! H_u^2-2{x_r}x\left[ n \right] - 2{y_r}y\left[ n \right] - 2{z_r}H_u. \hfill \\ 
	\end{gathered}
\end{equation*}
Thus ${\rm C}1.1$ can be written as 
\begin{equation*}\small
	\begin{gathered}
	{\rm C}1.2:p[n](2\Re \left[ {{\mathbf{\tilde d}}_{g,0}^T[n]{\mathbf{H}}_g^H[n]{{\mathbf{H}}_{\text{g}}}[n]{{{\mathbf{\tilde d}}}_g}[n]} \right]-\\
	~~~~~~~~~~~~~~~~~~~~{{\tilde{\bf{d}}}_{g,0}}^T[n]{\bf{H}}_{g}^H[n]{{\bf{H}}_{g}}[n]{{\tilde {\bf{d}}}_{g,0}}[n]) \ge {S^{ - 1}}[n].
\end{gathered}
\end{equation*}
Then we can substitute all concave points to solvable ones finally. Accordingly, (\rm{P3.3}) can be formulated as 
\begin{equation*}\label{20}
	\begin{split}
		\left( {{\rm{P}}3.4} \right):&\mathop {\max }\limits_{~~~~\bf{Q},S,G, \hfill\atop
		~	 {\xi _1}, {\xi _2}, {\xi _3}, {\xi _4}\hfill} \frac{1}{N}\sum\limits_{n \in {\cal{N}}} {{\tilde{R}}\left( {S[n],G[n]} \right)} \hfill\\
		{\rm{s}}{\rm{.t}}{\rm{.}}~
		&{F_1}\left[ n \right] -{ {{{\tilde \xi }_1}\left[ n \right]} } \leqslant 0,{F_2}\left[ n \right] - { {{{\tilde \xi }_2}\left[ n \right]} } \leqslant 0,\forall n,\\
		&{{\xi _3}^{ - 2}\left[ n \right]} - d\left[ n \right] \leqslant 0,{{\xi _4}^{ - 2}\left[ n \right]}  + {{\tilde F}_2}\left[ n \right] \leqslant 0,\forall n,\\
		&{\rm C}1.2, {\rm C}2.1, \left( 3 \right)-\left( 4 \right), \left( 18 \right).\\
	\end{split}	
\end{equation*}
(\rm{P3.4}) is a standard convex optimization problem, and we can use CVX solver to solve it.
\begin{algorithm}[t]
	\renewcommand{\algorithmicrequire}{\textbf{Input:}}
	\renewcommand{\algorithmicensure}{\textbf{Output:}}
	\caption{An alternating algorithm for solving {(\rm{P0}})}\label{Algorithm1}
	\begin{algorithmic} [1]
    \STATE\textbf{Input:}${\mu _2}$, ${i_{\max }}$
    \STATE\textbf{Ontput:}$R$. 
	\STATE{Initialization:}
	~Set~$i = 0$, as iteration index, ${\mu _2}$ as the threshold and original points ${\Upsilon _0} = \left\{ {{{\mathbf{Q}}^{\left( 0 \right)}},{{\mathbf{P}}^{\left( 0 \right)}},{{\mathbf{\Gamma }}^{\left( 0 \right)}}} \right\}$, thus obtaining the ${R^{\left( 0 \right)}}$ by using (\ref{R0}) and generating a series of initial points 
	${\mathbf{S}}_0^{\left( 0 \right)},{\mathbf{T}}_0^{\left( 0 \right)},{\mathbf{\xi }}_{10}^{\left( 0 \right)},{\mathbf{\xi }}_{20}^{\left( 0 \right)},{\mathbf{h}}_{ru}^{\left( 0 \right)},{\mathbf{ d}}_{g,0}^{\left( 0 \right)}, {\mathbf{ d}}_{m,0}^{\left( 0 \right)}$.
	\STATE\textbf{~repeat:}
	\STATE~ With given ${{\mathbf{Q}}^{\left( i \right)}}$, ${{\mathbf{\Gamma }}^{\left( i \right)}}$, update ${{\mathbf{P}}^{\left( i \right)}}$ to ${{\mathbf{P}}^{\left( i+1 \right)}}$  by solving sub-problem (\rm{P1}).
	\STATE~ With given ${{\mathbf{Q}}^{\left( i \right)}}$, ${{\mathbf{P}}^{\left( {i + 1} \right)}}$ , update ${{\mathbf{\Gamma }}^{\left( i \right)}}$  to ${{\mathbf{\Gamma }}^{\left( i+1 \right)}}$  by solving sub-problem (\rm{P2.4}). 	
	\STATE~ With given ${{\mathbf{\Gamma }}^{\left( {i + 1} \right)}}$, ${{\mathbf{P}}^{\left( {i + 1} \right)}}$ , update ${{\mathbf{Q}}^{\left( i \right)}}$ to ${{\mathbf{Q}}^{\left( i+1 \right)}}$ by solving sub-problem (\rm{P3.4}).
	\STATE~ With given ${{\mathbf{Q}}^{\left( i+1\right)}}$, ${{\mathbf{P}}^{\left( i+1 \right)}}$, ${{\mathbf{\Gamma }}^{\left( i+1 \right)}}$, compute ${R^{\left( {j + 1} \right)}}$ and ${\mathbf{S}}_0^{\left({i+1}\right)}$, ${\mathbf{T}}_0^{\left( {i + 1} \right)}$, ${\mathbf{\xi }}_{10}^{\left( {i + 1} \right)}$, ${\mathbf{\xi }}_{20}^{\left( {i + 1} \right)}$, ${\mathbf{h}}_{ru}^{\left( {i + 1} \right)}$, ${\mathbf{d}}_{g,0}^{\left( {i + 1} \right)}$, ${\mathbf{d}}_{m,0}^{\left( {i + 1} \right)}$.
	\STATE~ {Update $ i \leftarrow i + 1\ $}.
	\STATE\textbf{~until}: $\left| {{R^{\left( {j + 1} \right)}} - {R^{\left( j \right)}}} \right| \leqslant {\mu _2}$ or $j = {j_{\max }}$. 	
\end{algorithmic}
\end{algorithm}
					
\subsection{Overall Algorithm}
					
The overall algorithm for solving (P0) is summarized in Algorithm 2. The ${\mu _2}$ is used to control the accuracy of convergence and ${j_{\max }}$ is the maximum number of iterations. In fact, solving sub-problem 2 and sub-problem 3 dominates the complexity of Algorithm 2. 
Specifically, the computational complexities of Algorithm 2 is approximately ${\cal O}(N{(L + 1)^{3.5}}{I_1}{I_2} + 8{N^{3.5}}{I_2})$, where $I_1$ is the number of iterations required for solving (P2.4) and $I_2$ is for (P0), respectively.
									
\section{Numerical Results}
	\label{Numerical Results}
In this part, we give the simulation results to verify the effectiveness of our proposed algorithm. Specifically, {\bf{``Proposed"}} refers to the scheme we proposed for jointly optimizing UAV trajectory, GN's power allocation and IRS passive beamforming. To study the impact of the deployment of IRS, we consider two setups. In particular, one is denoted by ``{\bf IRS-M}" corresponding to that the IRS is deployed at (251, 50, 5), i.e., nearby the jammer;  while the other is denoted by ``{\bf IRS-G}" corresponding to that the IRS is deployed at (201, 100, 5), i.e., nearby the GN. Moreover, we consider benchmark algorithm that {\bf{``w/o IRS"}}  refers to the case without the IRS.
The parameters are set following the suggestions and settings in \cite{3}, and are shown as follows: ${{\rm{{\bf{q}}}}_{0}} = \left( {0,0,100} \right)$, ${{\rm{{\bf{q}}}}_{N}} = \left( {400,200,100} \right)$, ${{\rm{{\bf{q}}}}_m} = \left( {250,50,0} \right)$, ${{\rm{{\bf{q}}}}_g} = \left( {200,100,0} \right)$, $H_{u} = 100$ m, ${V_{\max }} = 60$, ${\bar p} = 30$ dBm, ${p_{max}} = 31.76$ dBm, ${p_{m}} = 30$ dBm, $\rho  = {10^{ - 3}}$, ${\sigma ^2} =  - 169$ dBm/Hz, ${\Delta t} = 0.5$, ${\mu _1} {\rm{ = }}{\mu _2}{\rm{ = }}{10^{ - 3}}$, $T=10^3$. 
					
\begin{figure}
	\centering
	\subfigure[$D_m=0$] {\includegraphics[width=.35\textwidth]{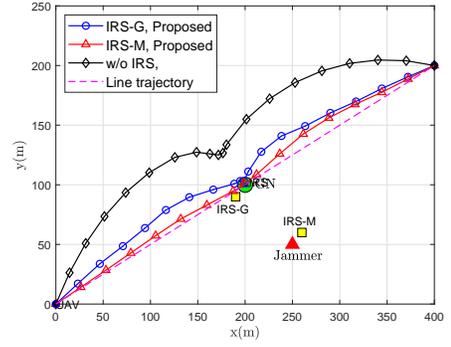}}
	\subfigure[$D_m=20$] {\includegraphics[width=.35\textwidth]{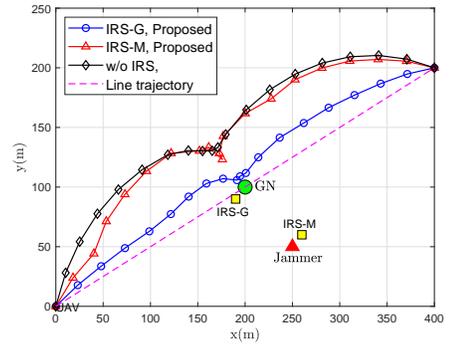}}
	\caption{UAV trajectories for different schemes.}
	\label{fig2tra}
	\vspace*{-5mm}
\end{figure}
					
Fig. \ref{fig2tra} shows the UAV's trajectories for different schemes when the number of IRS elements is $L=100$ and jammer's power is $p_m=30$ dBm. Fig. \ref{fig2tra}(a) shows the optimized UAV's trajectories when $D_m=0$. It is obviously shown that the trajectories in our proposed algorithm can significantly decrease the flying path length of the UAV compared to the case without IRS. Specifically, when deploying the IRS near the GN, the UAV keeps away from the jammer a bit first and finally hovers directly above the GN. While for the case when IRS is deployed near the jammer, it is observed that the UAV almost flies along the straight line from the start point to the end point. The reason is that the jamming signal power is drastically weakened by passive beamforming, which is roughly equivalent to the case that UAV flies under the low jamming level. Note that for the case ``w/o IRS", the UAV must keep away from the jammer farther and hovers for a while at a relative stable position so as to achieve an optimal system performance.

Fig. \ref{fig2tra}(b) shows the UAV's trajectories when $D_m=20$. It is observed that the trajectory for ``IRS-M"  becomes similar to that for ``w/o IRS". The reason is that when deploying the IRS nearby the jammer, the passive beamforming is quite sensitive to the location of the jammer, which is however incurs a large uncertainty and thus renders the passive beamforming focusing on mitigating the jamming signal  ineffective. In contrast, it can be observed that deploying the IRS nearby the GN is still helpful. This is because in this case the IRS mainly contributes to enhance the GN's signal received at the UAV but not reduce the jamming signal, which is thus not impacted by the uncertainty of the jammer's location..

\begin{figure}
	\centerline
	{\includegraphics
		[width=0.80\columnwidth]
		{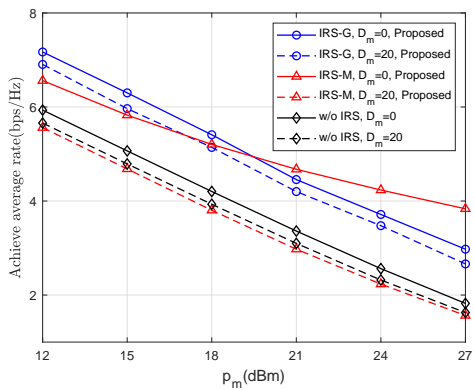}}
	\caption{\label{fig3pm}Achievable average rate for different schemes versus $p_m$.}
	\vspace{-5mm}
\end{figure}
	
Fig. \ref{fig3pm} shows the achievable average rate versus $p_m$ under the known and unknown jammer's location when $L=100$, respectively. For the case $D_m=0$, it can be observed that deploying the IRS nearby the GN and nearby the jammer both can increase the average uplink rate, for enhancing the information signal and reducing the jamming signal, respectively. Moreover, it is observed that ``IRS-G" outperforms ``IRS-M" first and then becomes less effective than the later. The reason is that when the jamming is under the low level, the SNR at the UAV is dominated by the noise, thus deploying the IRS nearby the GN for improving the information signal is more useful to increase the rate. However, as $p_m$ increases up to sufficiently large, the noise becomes ignorable and thus deploying the IRS nearby the jammer for jamming reduction becomes more effective. For the case $D_m=20$, it is observed that the achievable uplink rates for the three schemes all decreased and that for ``IRS-M" even becomes lower than that for ``w/o IRS". The reason is that the passive beamforming for IRS needs accurate location information to align the reflecting channels to the direct channel. Due to jammer's uncertain location, the jamming signal via IRS cann't be destructively combined with that from the direct channel for jamming reduction. Compared to the case ``IRS-M", ``IRS-G" is insensitive to the jammer's location and can still enhance the information-carrying signals. 

\begin{figure}
	\centerline
	{\includegraphics
		[width=0.80\columnwidth]
		{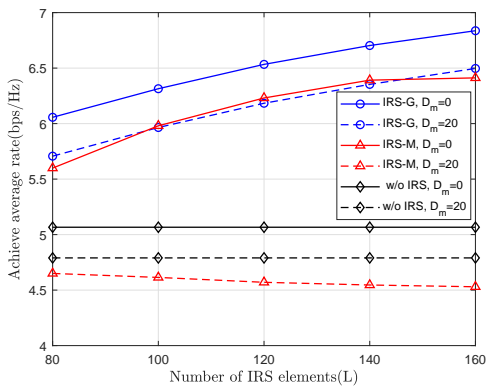}}
	\caption{\label{fig4L}Achievable average rate for different schemes versus $L$.}
	\vspace{-5mm}
\end{figure}

In Fig. \ref{fig4L}, the achievable average rate for different schemes versus the number of IRS elements are illustrated given $p_m=15$ dBm. It is observed when $D_m=0$, the achievable average rate for ``IRS-M" increases at the beginning and then tends to stability while ``IRS-G" keeps a steady growth rate. The reason is that as $L$ increases up to sufficiently large, the jamming signal in ``IRS-M" case is well reduced and thus the reception at the UAV is no more jamming-dominant, leading the lower performance gain by increasing the number of IRS elements. While for ``IRS-G", the reception at the UAV can substantially benefit from increasing $L$ because the IRS in this case mainly focuses on enhancing the information signal from GN. When $D_m=20$, it is observed that the average rate for ``IRS-G" is much higher than that for the other two and still benefits from increasing the number of IRS elements. However, the rate for ``IRS-M" is lower than that for ``w/o IRS" and even decreases as increasing $L$. This is because for ``IRS-G", the passive beamforming gain is insensitive the the location of the jammer and a larger number of IRS elements is always favorable to improve the rate; while for ``IRS-M", the passive beamforming can no more reduce the jamming signal effectively due to the uncertain location of the jammer, and even helps enhance the jamming singal at the UAV statistically.

\section{Conclusions}\label{Conclusions}
					
In this paper, we study the UAV uplink transmission assisted by IRS in the presence of a jammer with imperfect location information. By considering the GN's power allocation, IRS passive beamforming, and UAV trajectory, an alternating optimization based algorithm is proposed to solve the problem by exploiting the SDR, SDA and S-procedure methods. Simulation results were shown to verify the performance of our proposed algorithm, by considering two setups of the IRS deployment for further exploiting the effect of IRS. Specifically, it is observed that by deploying the IRS nearby the GN, the proposed joint design can always improve the uplink transmission regardless of the knowledge of jammer's location; however, by deploying the IRS nearby the jammer, the proposed design is effective only when the jammer's location is perfectly known.

\bibliographystyle{IEEEtran}

\begin{thebibliography}{1}
\balance
						
\bibitem{1}
Y. Zeng, Q. Wu and R. Zhang, ``Accessing From the Sky: A Tutorial on UAV Communications for 5G and Beyond," \emph{Proceedings of the IEEE}, vol. 107, no. 12, pp. 2327$-$2375, Dec. 2019.	
\bibitem{2}
J. Xu, Y. Zeng and R. Zhang, ``UAV-Enabled Wireless Power Transfer: Trajectory Design and Energy Optimization," \emph{IEEE Trans. Wireless Commun.}, vol. 17, no. 8, pp. 5092$-$5106, Aug. 2018. 
\bibitem{3}
Y. Wu, W. Fan, W. Yang, X. Sun and X. Guan, ``Robust Trajectory and Communication Design for Multi-UAV Enabled Wireless Networks in the Presence of Jammers," \emph{IEEE Access}, vol. 8, pp. 2893$-$2905, 2020.
\bibitem{5}
Q. Wu and R. Zhang, ``Intelligent reflecting surface enhanced wireless network via joint active and passive beamforming,” \emph{IEEE J. Sel. Areas Commun.}, vol. 38, no. 8, pp. 1719$-$1734, Aug. 2020.
\bibitem{huang1}
C. Yuen, C. Huang, I. F. Akyildiz, M. Di Renzo and M. Debbah, ``Guest Editorial: Intelligent Surfaces for 5G and Beyond," \emph{IEEE Wireless Commun.}, vol. 28, no. 6, pp. 70$-$71, Dec. 2021.
\bibitem{cog2}
X. Guan, Q. Wu, and R. Zhang, ``Joint power control and passive beamforming in IRS-assisted spectrum sharing,”\emph{IEEE Commun. Lett.}, vol. 24, no. 7, pp. 1553$-$1557, Jul. 2020.
\bibitem{8}
M. H. Khoshafa, T. M. N. Ngatched and M. H. Ahmed, ``Reconfigurable Intelligent Surfaces-Aided Physical Layer Security Enhancement in D2D Underlay Communications," \emph{IEEE Commun. Lett.}, vol. 25, no. 5, pp. 1443$-$1447, May. 2021.
\bibitem{11}
X. Guan, Q. Wu, and R. Zhang, ``Intelligent reflecting surface assisted secrecy communication: Is artificial noise helpful or not?" \emph{IEEE Wireless Commun. Lett.}, vol. 9, no. 6, pp. 778$-$782, Jun. 2020.
\bibitem{12}
Q. Wu, S. Zhang, B. Zheng, C. You, and R. Zhang, ``Intelligent Reflecting Surface-Aided Wireless Communications: A Tutorial," \emph{IEEE Trans. Commun.}, vol. 69, no. 5, pp. 3313$-$3351, May. 2021.
\bibitem{17}
X. Pang, N. Zhao, J. Tang, C. Wu, D. Niyato and K. -K. Wong, ``IRS-Assisted Secure UAV Transmission via Joint Trajectory and Beamforming Design,"\emph{IEEE Trans. Commun.}, doi: 10.1109/TCOMM.2021.3136563.
\bibitem{ICC}
J. Fang, Z. Yang, N. Anjum, Y. Hu, H. Asgari and M. Shikh-Bahaei, ``Secure Intelligent Reflecting Surface Assisted UAV Communication Networks,"\emph{in Proc. IEEE ICC Workshops.}, 2021, pp. 1$-$6.
\bibitem{WCL}
S. Fang, G. Chen and Y. Li, ``Joint Optimization for Secure Intelligent Reflecting Surface Assisted UAV Networks," \emph{IEEE Wireless Commun. Lett.}, vol. 10, no. 2, pp. 276$-$280, Feb. 2021, 
\bibitem{21}
Z. Ji, W. Yang, X. Guan, X. Zhao, G. Li, and Q. Wu ``Trajectory and Transmit Power Optimization for IRS-Assisted UAV Communication under Malicious Jamming,” \emph{arXiv preprint }, arXiv: 2201.05271, 2022.
\bibitem{20}
S. Li, B. Duo, M. D. Renzo, M. Tao and X. Yuan,  ``Robust Secure UAV Communications With the Aid of Reconfigurable Intelligent Surfaces,”\emph{IEEE Trans. Wireless Commun.}, vol. 20, no. 10, pp. 6402$-$6417, Oct. 2021.
\bibitem{24}
X. Lu, W. Yang, X. Guan, Q. Wu and Y. Cai, ``Robust and Secure Beamforming for Intelligent Reflecting Surface Aided mmWave MISO Systems,” \emph{IEEE Wireless Commun. Lett.} , vol. 9, no. 12, pp. 2068$-$2072, Dec. 2020.
\end{thebibliography}

\end{document}